\documentclass[letterpaper, 10 pt, conference]{ieeeconf}  % Comment this line out
\usepackage{amssymb,amsfonts,amsmath,amstext,mathrsfs}
\usepackage{latexsym}
\usepackage{epsfig,psfrag,color,graphics,epsf}
\usepackage{enumerate,cite}

\newcommand{\bee}{\begin{eqnarray}}
\newcommand{\eee}{\end{eqnarray}}

\newtheorem{definition}{\textbf{Definition}}
\newtheorem{corollary}{\textbf{Corollary}}
\newtheorem{lemma}{\textbf{Lemma}}
\newtheorem{theorem}{\textbf{Theorem}}

\newtheorem{remark}{\textbf{Remark}}

\newcommand{\spp}{\hspace{5mm}}

\newcommand{\cX}{\mathcal{X}}

\newcommand{\cY}{\mathcal{Y}}

\newcommand{\cZ}{\mathcal{Z}}

\newcommand{\cW}{\mathcal{W}}

\newcommand{\cC}{\mathcal{C}}

\newcommand{\cT}{\mathcal{T}}

\newcommand{\hw}{\hat{w}}

\DeclareMathAlphabet{\matheuf}{U}{euf}{m}{n}
\newcommand{\eufC}{\mathscr{C}}
\newcommand{\eufR}{\mathscr{R}}

\newcommand{\qq}{\vspace{0.2cm}}
\IEEEoverridecommandlockouts                              % This command is only
\title{\LARGE \bf
Cognitive Interference Channels with Confidential Messages }

\author{Yingbin Liang, Anelia Somekh-Baruch, H. Vincent Poor, \\
Shlomo Shamai (Shitz), and Sergio Verd\'{u}% <-this % stops a space
\thanks{The work of Y. Liang and H. V. Poor was supported by the National
Science Foundation under Grants ANI-03-38807 and CNS-06-25637. The
work of A. Somekh-Baruch was supported by a Marie Curie Outgoing
International Fellowship within the $6$th European Community
Framework Programme. The work of S. Shamai and S. Verd\'{u} was
supported by the US-Israel Binational Science Foundation.
}% <-this % stops a space
\thanks{Yingbin Liang, Anelia Somekh-Baruch, H. Vincent Poor,
and Sergio Verd\'{u} are with the Department of Electrical
Engineering, Princeton University, E-Quad, Olden Street, Princeton,
NJ 08544, USA {\tt\small \{yingbinl,anelia,poor,verdu\}@princeton.edu}}%
\thanks{Shlomo Shamai (Shitz) is
with the Department of Electrical Engineering, Technion-Israel
Institute of Technology, Technion City, Haifa 32000, Israel
{\tt\small sshlomo@ee.technion.ac.il}}%
}

\begin{document}

\maketitle \thispagestyle{empty} \pagestyle{empty}

%%%%%%%%%%%%%%%%%%%%%%%%%%%%%%%%%%%%%%%%%%%%%%%%%%%%%%%%%%%%%%%%%%%%%%%%%%%%%%%%
\begin{abstract}
The cognitive interference channel with confidential messages is
studied. Similarly to the classical two-user interference channel,
the cognitive interference channel consists of two transmitters
whose signals interfere at the two receivers. It is assumed that
there is a common message source (message 1) known to both
transmitters, and an additional independent message source
(message 2) known only to the cognitive transmitter (transmitter
$2$). The cognitive receiver (receiver $2$) needs to decode both
messages, while the non-cognitive receiver (receiver $1$) should
decode only the common message. Furthermore, message $2$ is
assumed to be a confidential message which needs to be kept as
secret as possible from receiver $1$, which is viewed as an
eavesdropper with regard to message $2$. The level of secrecy is
measured by the equivocation rate. A single-letter expression for
the capacity-equivocation region of the discrete memoryless
cognitive interference channel is established and is further
explicitly derived for the Gaussian case. Moreover,
particularizing the capacity-equivocation region to the case
without a secrecy constraint, establishes a new capacity theorem
for a class of interference channels, by providing a converse
theorem.
\end{abstract}

%%%%%%%%%%%%%%%%%%%%%%%%%%%%%%%%%%%%%%%%%%%%%%%%%%%%%%%%%%%%%%%%%%%%%%%%%%%%%%%%
\section{Introduction}

Interference channels model basic wireless networks, in which
communication signals intended for one receiver cause interference
at other receivers. %Such an issue has been studied from the
%information-theoretic viewpoint in terms of rate regions and
%coding strategies for the interference channel.
Although the capacity region and the best coding schemes for the
interference channel remain unknown, much progress has been made
toward understanding this channel (see, e.g.,
\cite{Carl78,Han81,Chong06,Kramer06,Etkin07,Telatar07} and the
references therein).

Interference not only affects communication rates, but also raises
security issues. This is because information on the original source
can be extracted by other nodes that are not the intended
destination (eavesdroppers) due to interference at these nodes. In
certain situations it is desirable to minimize the leak of
information to eavesdroppers. It is also important to evaluate the
secrecy level of confidential information (which was defined in
\cite{Wyner75} as the equivocation rate for the single-user wire-tap
channel) for the interference channel, and to study the reliable
communication rates under a given certain level of secrecy
constraint. This is what motivates the problem that we address in
this paper.

\begin{figure*}[tbhp]
\begin{center}
\includegraphics[width=14cm]{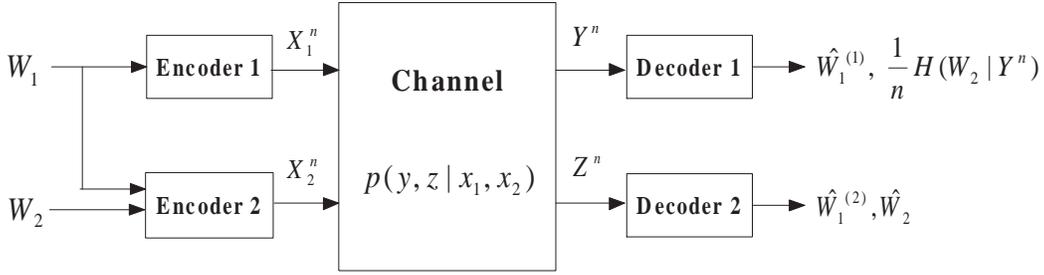}
\caption{Cognitive interference channel with confidential
messages} \label{fig:ICframe}
\end{center}
\end{figure*}

We study the cognitive interference channel, with two transmitters
and two receivers (see Fig.~\ref{fig:ICframe}). Transmitter 1
knows only message 1, and transmitter 2 (the cognitive
transmitter) knows both messages 1 and 2. Receiver 1 needs to
decode only message 1 and receiver 2 needs to decode both messages
1 and 2. Furthermore, we also address a secrecy constraint, and
assume that transmitter 2 wishes to protect message 2
(confidential message) from being known by receiver 1 (an
eavesdropper) due to interference. We refer to this model as {\em
the cognitive interference channel with confidential messages}.

The channel studied in this paper can model a cognitive radio (see
\cite{Devroye-Mitran-Tarokh-IT06,Devroye-Mitran-Tarokh-CommunicationMagazine06,Wu06,Jovicic06}),
which is a device that is introduced to wireless networks in order
to improve overall performance by utilizing unused spectral
resources. In the model we study, the cognitive radio is modelled
by transmitter 2 which helps transmitter 1 (the primary
transmitter) transmit its message. Moreover, the cognitive
transmitter also transmits its own message, which should be kept
confidential with respect to the non-cognitive receiver (receiver
1).

In this paper, we establish the capacity-equivocation region for the
cognitive interference channel with confidential messages, which
characterizes the trade-off between the achievable communication
rates and the secrecy levels at the eavesdropper. We further derive
the capacity-equivocation region for the Gaussian case. For the case
without the secrecy constraint, the capacity-equivocation region
reduces to the capacity region of the cognitive interference
channel. This establishes a new capacity theorem for a class of
interference channels, by providing a converse theorem.

We note that the cognitive interference channel (without secrecy
constraints) was studied in \cite[Theorem~5]{Jiang06}, where an
achievable rate region (inner bound on the capacity region) is
given. An achievable error exponent for this channel was studied in
\cite{Zhong07}. In this paper, we provide an outer bound on the
capacity region that matches the inner bound given in
\cite[Theorem~5]{Jiang06} and hence establish the capacity region
for this channel. We also note that the channel we study is
different from the channel model studied in
\cite{Wu06,Jovicic06,Maric07,Maric07_2} in that receiver 2 is
required to decode both messages 1 and 2. Furthermore, we also
address the secrecy constraint which was not considered in
\cite{Jiang06,Wu06,Jovicic06,Maric07,Maric07_2,Zhong07}. We finally
note that the model we study is different from the interference
channel model with secrecy constraints studied in \cite{Liu07}, in
that it assumes a cognitive transmitter that knows the other
transmitter's message and a cognitive receiver that decodes both
messages.

The paper is organized as follows. In Section \ref{sec:model} we
introduce the model of the cognitive interference channel with
confidential messages. Section \ref{sec:mainresults} presents the
capacity-equivocation region, and Section \ref{sec:gauss} applies
our results to the Gaussian case. Section \ref{sec:implication}
particularizes our results to the cognitive interference channel
without secrecy constraints.

\section{Channel Model}\label{sec:model}

\begin{definition}\label{def:mac}
A discrete memoryless cognitive interference channel consists of two
finite channel input alphabets $\cX_1$ and $\cX_2$, two finite
channel output alphabets $\cY$ and $\cZ$, and a transition
probability distribution $p(y,z|x_1,x_2)$ (see
Fig.~\ref{fig:ICframe}), where $x_1 \in \cX_1$ and $ x_2 \in \cX_2$
are channel inputs from transmitters 1 and 2, respectively, and $y
\in \cY$ and $z \in \cZ$ are channel outputs at receivers 1 and 2,
respectively.
\end{definition}

Over this channel, transmitters 1 and 2 jointly send one message
denoted by $W_1$ to receivers 1 and 2, and transmitter 2 sends one
message denoted by $W_2$ to receiver 2 and wants to keep this
message as secret as possible from receiver 1. Hence the message
$W_2$ is referred to the confidential message with respect to
receiver 1.

\begin{definition}
A $\left(2^{nR_1},2^{nR_2}, n \right)$ code for the cognitive
interference channel consists of the following:
\begin{list}{$\bullet$}{\topsep=0ex \leftmargin=3mm
\rightmargin=2mm \itemsep=1mm}

\item Two message sets: $\cW_k=\{1,2,\ldots,2^{nR_k}\}$ for
$k=1,2$;

\item Two messages: $W_1$ and $W_2$ are independent random
variables that are uniformly distributed over $\cW_1$ and $\cW_2$,
respectively;

\item Two encoders: one deterministic encoder $f_1$: $\cW_1
\rightarrow \cX_1^n$, which maps each message $w_1 \in \cW_1$ to a
codeword $x_1^n \in \cX_1^n$; and one stochastic
encoder\footnote{We note that the stochastic encoder $f_2$ defines
a transition probability distribution $f_2(x_2^n|w_1,w_2)$. In
fact, $f_2$ can be equivalently represented by a deterministic
mapping $\cW_1 \times \cW_2 \times \cT \rightarrow \cX_2^n$, which
maps $(w_1,w_2,t) \in \cW_1 \times \cW_2 \times \cT$ to a codeword
$x_2^n \in \cX_2^n$, where $t$ is a realization of a randomizing
variable $T$ that is independent of $(W_1,W_2)$. The distribution
of $T$ is part of the encoding strategy of transmitter 2. We
assume this distribution is known at both receivers, but the
realization of $T$ is not known at either receiver.} $f_2$: $\cW_1
\times \cW_2 \rightarrow \cX_2^n$, which maps each message pair
$(w_1,w_2) \in \cW_1 \times \cW_2 $ to a codeword $x_2^n \in
\cX_2^n$;

\item Two decoders: one is $g_1$: $\cY^n \rightarrow \cW_1 $,
which maps a received sequence $y^n$ to a message $\hw^{(1)}_1 \in
\cW_1$; and the other is $g_2$: $\cZ^n \rightarrow \cW_1\times \cW_2
$, which maps a received sequence $z^n$ to a message pair
$(\hw^{(2)}_1,w_2) \in \cW_1\times \cW_2$.
\end{list}
\end{definition}

For a given code, we define the probability of error and the secrecy
level of the confidential message $W_2$. The probability of error
when the message pair $(w_1,w_2)$ is sent is defined as
\begin{equation}
P_e^{(n)}(w_1,w_2)=Pr\left\{ (\hw^{(1)}_1,\hw^{(2)}_1,\hw_2)\neq
(w_1,w_1,w_2)\right\}
\end{equation}
and the average block probability of error is
\begin{equation}
P_e^{(n)} = \frac{1}{2^{n(R_1+R_2)}} \sum_{w_1=1}^{2^{nR_1}}
\sum_{w_2=1}^{2^{nR_2}}P_e^{(n)}(w_1,w_2).
\end{equation}

The secrecy level of the message $W_2$ at receiver 1 is defined by
%%the equivocation rate
\begin{equation}\label{eq:equiv1}
R_{2e}^{(n)}=\frac{1}{n} H(W_2|Y^n).
\end{equation}
%%which is denoted by $R_{2e}$.

A rate-equivocation triple $(R_1,R_2,R_{2e})$ is said to be {\em
achievable} if there exists a sequence of
$\left(2^{nR_1},2^{nR_2}, n \right)$ codes for $n \geq 1$ with the
average error probability
\[P_e^{(n)}\rightarrow 0\]
as $n\rightarrow \infty$ and with the equivocation rate, $R_{2e}$,
satisfying
\begin{equation*}
R_{2e} \leq \liminf_{n\rightarrow \infty} R_{2e}^{(n)}. %\frac{1}{n}H(W_2|Y^n).
\end{equation*}

\begin{definition}
The {\em capacity-equivocation region} $\eufC$ is the closure of
the union of all achievable rate-equivocation triples
$(R_1,R_2,R_{2e})$.
\end{definition}

\begin{definition}\label{def:sec}
The {\em secrecy capacity region}, $\cC_s$, is defined by
\begin{equation}
\cC_s=\left\{(R_1,R_2): (R_1,R_2,R_2) \in \eufC \right\},
\end{equation}
that is, the region that includes all achievable rate-pairs
$(R_1,R_2)$ such that perfect secrecy is achieved for the message
$W_2$.
\end{definition}

\section{Main Results}\label{sec:mainresults}

We first provide an achievable rate-equivocation region for the
cognitive interference channel in the following lemma.
\begin{lemma}\label{lemma:achieve}
The following region is achievable for the cognitive interference
channel with confidential messages:
\begin{equation}\label{eq:achieve}
\begin{split}
& \eufR = \bigcup_{\displaystyle p(u,x_1)p(x_2|u,x_1)
p(y,z|x_1,x_2)} \\
& \left\{
\begin{array}{l}
(R_1,R_2,R_{2e}): \\
R_1 \ge 0, R_2 \ge 0, R_{2e} \ge 0, R_{21} \ge 0, R_{22} \ge 0 \\
R_2=R_{21}+R_{22} \\
R_1+R_{21} \leq I(U,X_1;Y) \\
R_{22} \leq I(X_2;Z|U,X_1) \\
R_{21}+R_{22} \leq I(U,X_2;Z|X_1) \\
R_1+R_{21}+R_{22} \leq I(U,X_1,X_2;Z) \\
R_{2e} \leq R_{22} \\
R_{2e} \leq I(X_2;Z|U,X_1)-I(X_2;Y|U,X_1) \\
R_{21}+R_{2e} \leq I(U,X_2;Z|X_1)-I(X_2;Y|U,X_1) \\
R_1+R_{21}+R_{2e} \leq I(U,X_1,X_2;Z)-I(X_2;Y|U,X_1)
\end{array} \right\}
\end{split}
\end{equation}
\end{lemma}
\begin{proof}(outline)
We briefly outline the idea of the achievable scheme in the
following. The details of the proof and the computation of the
equivocation rate are omitted and can be found in \cite{Liang07}.
The message $W_2$ is split into two components, $W_{21}$ and
$W_{22}$, with rates indicated by $R_{21}$ and $R_{22}$,
respectively, in \eqref{eq:achieve}. Receiver 1 decodes both $W_1$
and $W_{21}$, and receiver 2 decodes $W_1$, $W_{21}$ and $W_{22}$.
Since $W_{21}$ is decoded and fully known at receiver 1, $W_{21}$
does not contribute to the secrecy level of $W_2$ at receiver 1
(the eavesdropper). Hence, only $W_{22}$ may be hidden from the
eavesdropper.
\end{proof}

We now present our main result in the following theorem.
\begin{theorem}\label{th:capaequi}
For the cognitive interference channel with confidential messages,
the capacity-equivocation region is given by
\begin{equation}\label{eq:capaequi}
\begin{split}
& \eufC = \bigcup_{\displaystyle p(u,x_1,v)p(x_2|v)
p(y,z|x_1,x_2)} \\
& \left\{
\begin{array}{l}
(R_1,R_2,R_{2e}): \\
R_1 \ge 0, R_2 \ge 0, R_{2e} \ge 0 \\
R_1 \leq I(U,X_1;Y) \\
R_2 \leq I(U,V;Z|X_1) \\
R_1+R_2 \leq \min\{ I(U,X_1;Y), I(U,X_1;Z) \} \\
\hspace{1.7cm}+I(V;Z|U,X_1) \\
R_{2,e} \leq R_2 \\
R_{2,e} \leq I(V;Z|U,X_1)-I(V;Y|U,X_1) \\
R_1+R_{2,e} \leq I(U,V,X_1;Z)-I(V;Y|U,X_1)
\end{array} \right\}
\end{split}
\end{equation}
\end{theorem}
\vspace{0.2cm}
\begin{proof}(outline)
To establish the achievability part of Theorem \ref{th:capaequi},
we first note that if we define a random variable $V$ that
satisfies the Markov chain condition $(X_1,U) \leftrightarrow V
\leftrightarrow X_2$, and change $X_2$ to be $V$ in $\eufR$ given
in Lemma \ref{lemma:achieve}, the resulting region is also
achievable. This follows by prefixing one discrete memoryless
channel with the input $V$ and the transition probability
$p(x_2|v)$ to transmitter 2 (similarly to \cite[Lemma
4]{Csiszar78}). For this new achievable region, we apply {\em
Fourier-Motzkin elimination} (see, e.g.,~\cite{LallNotes}) to
eliminate $R_{21}$ and $R_{22}$ from the bounds and then obtain
the region $\eufC$ in Theorem \ref{th:capaequi}.

The proof of the converse part of Theorem \ref{th:capaequi} is
omitted and can be found in \cite{Liang07}.
\end{proof}

The capacity-equivocation region provided in Theorem
\ref{th:capaequi} in \eqref{eq:capaequi} can also be written in
the following equivalent form.
\begin{corollary}\label{cor:capaequi1}
The following region is equivalent to the region in Theorem
\ref{th:capaequi} in \eqref{eq:capaequi} and provides a simpler
form of the capacity-equivocation region for the cognitive
interference channel with confidential messages:
\begin{equation}\label{eq:capaequi1}
\begin{split}
& \eufC = \bigcup_{\displaystyle p(u,x_1,v)p(x_2|v)
p(y,z|x_1,x_2)} \\
& \left\{
\begin{array}{l}
(R_1,R_2,R_{2e}): \\
R_1 \ge 0, R_2 \ge 0, R_{2e} \ge 0 \\
R_1 \leq \min\{ I(U,X_1;Y), I(U,X_1;Z) \} \\
R_2 \leq I(U,V;Z|X_1) \\
R_1+R_2 \leq \min\{ I(U,X_1;Y), I(U,X_1;Z) \} \\
\hspace{1.7cm}+I(V;Z|U,X_1) \\
R_{2,e} \leq R_2 \\
R_{2,e} \leq I(V;Z|U,X_1)-I(V;Y|U,X_1)
\end{array} \right\}
\end{split}
\end{equation}
\end{corollary}
%We observe that the cognitive interference channel with confidential
%messages reduces to the broadcast channel with confidential
%messages studied in \cite{Csiszar78} if $X_1$ is set to be null.
\qq

\begin{remark}
The capacity-equivocation region of the cognitive interference
channel with confidential messages given in \eqref{eq:capaequi1}
reduces to the capacity-equivocation region of the broadcast
channel with confidential messages given in
\cite[Theorem~1]{Csiszar78} when setting $X_1=\phi$.
\end{remark}

For the case of perfect secrecy, we obtain the following secrecy
capacity region based on Corollary \ref{cor:capaequi1}.
\begin{corollary}\label{cor:capasec}
The secrecy capacity region of the cognitive interference channel
with confidential messages is given by:
\begin{equation}\label{eq:capasec}
\begin{split}
& \cC_s = \bigcup_{\displaystyle p(u,x_1,v)p(x_2|v)
p(y,z|x_1,x_2)} \\
& \left\{
\begin{array}{l}
(R_1,R_2): \\
R_1 \ge 0, R_2 \ge 0 \\
R_1 \leq \min\{ I(U,X_1;Y), I(U,X_1;Z) \} \\
R_2 \leq I(V;Z|U,X_1)-I(V;Y|U,X_1)
\end{array} \right\}
\end{split}
\end{equation}
\end{corollary}
\vspace{0.2cm}

We next present the capacity-equivocation region for two classes
of degraded cognitive interference channels, which will be useful
when we study the Gaussian case.
\begin{corollary}\label{cor:capaequi degraded1}
If the cognitive interference channel satisfies the following
degradedness condition \bee p(y,z|x_1,x_2)=p(y|x_1,x_2)p(z|y,x_1),
\label{eq:degradedness condition1} \eee then the
capacity-equivocation region is given by
\begin{equation}\label{eq:capaequi degraded1}
\begin{split}
& \eufC_{deg1} = \bigcup_{\displaystyle p(x_1,x_2)
p(y,z|x_1,x_2)} \\
& \left\{
\begin{array}{l}
(R_1,R_2,0): \\
R_1 \ge 0, R_2 \ge 0 \\%, R_{2,e}= 0 \\
R_2 \leq I(X_2;Z|X_1) \\
R_1+R_2 \leq \min\{I(X_1,X_2;Y),I(X_1,X_2;Z)\}
\end{array} \right\}
\end{split}
\end{equation}
\end{corollary}
\begin{proof}(outline)
The achievability follows from \eqref{eq:capaequi} given in
Theorem \ref{th:capaequi} by setting $U=V=X_2$. The proof of the
converse part is omitted and can be found in \cite{Liang07}.
\end{proof}

\begin{corollary}\label{cor:capaequi degraded2}
If the cognitive interference channel satisfies the following
degradedness condition \bee
p(y,z|x_1,x_2)=p(z|x_1,x_2)p(y|z,x_1),\label{eq:degradedness
condition2} \eee then the capacity-equivocation region is given by
\begin{equation}\label{eq:capaequi degraded2}
\begin{split}
& \eufC_{deg2} = \bigcup_{\displaystyle p(u,x_1,x_2)
p(y,z|x_1,x_2)} \\
& \left\{
\begin{array}{l}
(R_1,R_2,R_{2e}): \\
R_1 \ge 0, R_2 \ge 0, R_{2,e}\geq 0 \\
R_1 \leq \min\left\{I(U,X_1;Y),I(U,X_1;Z)\right\} \\
R_2 \leq I(X_2;Z|U,X_1) \\
R_{2,e} \leq R_2 \\
R_{2,e} \leq I(X_2;Z|U,X_1)-I(X_2;Y|U,X_1)
\end{array} \right\}
\end{split}
\end{equation}
\end{corollary}
\qq
\begin{proof}(outline)
The achievability follows from (\ref{eq:capaequi1}) by setting
$V=X_2$ and observing that the sum-rate bound in
(\ref{eq:capaequi1}) is equal to the sum of the two bounds on the
individual rates in (\ref{eq:capaequi degraded2}) and that
$I(X_2;Z|U,X_1)\leq I(U,X_2;Z|X_1)$. The proof of the converse
part is omitted and can be found in \cite{Liang07}.
\end{proof}

Corollary \ref{cor:capaequi degraded1}, (and similarly, Corollary
\ref{cor:capaequi degraded2}) continues to hold also for a
stochastically degraded channel, i.e., a channel $p(y,z|x_1,x_2)$
whose conditional marginal distributions $p(y|x_1,x_2)$ and
$p(z|x_1,x_2)$ are the same as those of a channel satisfying the
degradedness condition \eqref{eq:degradedness condition1} (and
correspondingly \eqref{eq:degradedness condition2} for Corollary
\ref{cor:capaequi degraded2}).

We note that while achieving the capacity-equivocation region for
the general cognitive interference channel with confidential
messages requires application of a rate splitting scheme (described
in the proof for Lemma \ref{lemma:achieve}), it is unnecessary in
the degraded channel cases (\ref{eq:degradedness condition2}),
(\ref{eq:degradedness condition1}).

\section{Gaussian Cognitive Interference Channel with Confidential Messages}\label{sec:gauss}

In this section, we consider the Gaussian cognitive interference
channel. The channel outputs at receivers 1 and 2 at time instant
$i$ are given, respectively, by
\begin{equation}\label{eq:Gaussian IC equations}
\begin{split}
Y_i&=X_{1,i}+aX_{2,i}+N_{1,i} \\
Z_i&=bX_{1,i}+X_{2,i}+N_{2,i}
\end{split}
\end{equation}
where $\{N_{1,i}\}_{i=1}^{\infty}$ and
$\{N_{2,i}\}_{i=1}^{\infty}$ are independent i.i.d.\ Gaussian
processes, and $a$ and $b$ are real constants. Without loss of
generality, we set $Var(N_{1,i})=Var(N_{2,i})=1$. We assume that
the transmitters are subject to the following power constraints
\begin{equation}\label{eq:Gaussian power constraints}
\frac{1}{n}\sum_{i=1}^n X_{1,i}^2\leq P_1 \spp \text{and} \spp
\frac{1}{n}\sum_{i=1}^n X_{2,i}^2\leq P_2.
\end{equation}
We consider the cases with $|a|\geq 1$ and $|a| < 1$, separately.
For the case when $|a|\geq 1$, we have the following theorem on the
capacity-equivocation region.
\begin{theorem}\label{th:gausscapaequi1}
For the Gaussian cognitive interference channel with confidential
messages, if $|a|\geq 1$, then the capacity-equivocation region is
given by
\begin{equation}\label{eq:gausscapaequi1}
\begin{split}
& \eufC = \bigcup_{-1 \leq \rho \leq 1} \\
& \left\{
\begin{array}{l}
(R_1,R_2,0): \\
R_1 \ge 0, R_2 \ge 0 \\
R_2 \leq \frac{1}{2}\log \left(1+(1-\rho^2) P_2 \right) \\
R_1+R_2 \leq \frac{1}{2}\log \left(1+b^2P_1+P_2+2b \rho\sqrt{P_1P_2} \right) \\
R_1+R_2 \leq \frac{1}{2}\log \left(1+P_1+a^2P_2+2a \rho\sqrt{P_1P_2} \right) \\
%R_{2e}=0 \\
\end{array} \right\}
\end{split}
\end{equation}
\end{theorem}

From Theorem \ref{th:gausscapaequi1}, we observe that no secrecy can
be achieved whenever $|a|\geq 1$, i.e., $R_{2e}=0$. This is because
when $|a|\geq 1$ and $X_1$ is given, receiver 2's output $Z$ is
degraded with regard to receiver 1's output $Y$, and hence receiver
1 can obtain any information that receiver 2 obtains.

\begin{proof}(outline)
The achievability follows from \eqref{eq:capaequi degraded1} given
in Corollary \ref{cor:capaequi degraded1} by computing the mutual
information terms with $(X_1,X_2)$ that are zero-mean jointly
Gaussian with $E[X_1^2]=P_1$, $E[X_2^2]=P_2$, and
$E[X_1X_2]=\rho\sqrt{P_1P_2}$.

The converse follows by applying the bounds in the converse proof
for Theorem \ref{th:capaequi} (see \cite{Liang07}) to the Gaussian
case. The power constraints \eqref{eq:Gaussian power constraints}
translate to upper bounds on the second moments of $X_1$ and
$X_2$, i.e., $Var(X_1)\leq P_1$ and $Var(X_2)\leq P_2$. The proof
also applies the degradedness property in this case, i.e., $Z$ is
degraded with regard to $Y$ if $X_1$ is given. The details of the
proof are omitted and can be found in \cite{Liang07}.
\end{proof}

\begin{figure}
\begin{center}
\psfrag{R1}[][][0.7]{$R_1$} \psfrag{R2}[][][0.7]{$R_2$}
\psfrag{a=1}[][][0.7]{$a=1$} \psfrag{a=2}[][][0.7]{$a=2$}
\psfrag{a>=3}[][][0.7]{$a\geq 3$}
\includegraphics[width=\linewidth]{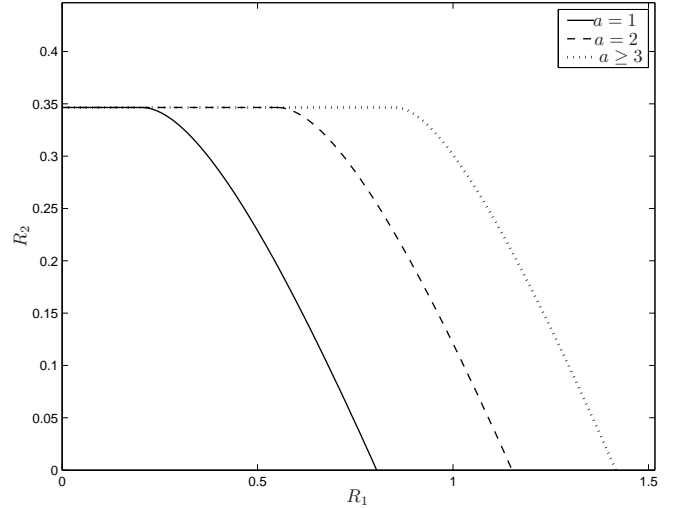}
\end{center}
\caption{The capacity region of the Gaussian cognitive
interference channel for $P_1=P_2=1$, $b=3$, and
$a>1$.}\label{fig:high interference Gaussian IC}
\end{figure}

In Figure \ref{fig:high interference Gaussian IC}, the capacity
region of the Gaussian cognitive interference channel is shown for
$P_1=P_2=1$, $b=3$, and $a=1,2,3$. In fact, in this case, the
capacity region is the same for all $a\geq 3$ as one can see in
(\ref{eq:gausscapaequi1}). This is because for the chosen
parameters $P_1=P_2=1$ and $b=3$, if $a\geq 3$, receiver 1 always
decodes $W_1$ if receiver 2 decodes this message. Hence receiver 2
is the bottleneck receiver.

We now show the capacity-equivocation region when $|a| < 1$.
\begin{theorem}\label{th:gausscapaequi2}
For the Gaussian cognitive interference channel with confidential
messages, if $|a| < 1$, the capacity-equivocation region is given by
\begin{equation}\label{eq:gausscapaequi2_deg}
\begin{split}
& \eufC = \bigcup_{-1 \leq \rho \leq 1, 0\leq\beta\leq 1} \\
& \left\{
\begin{array}{l}
(R_1,R_2,R_{2e}): \\
R_1 \ge 0, R_2 \ge 0, R_{2,e} \ge 0\\
R_1 \leq \frac{1}{2}\log \left(1+\frac{P_1+\rho^2 a^2 P_2+2\rho a \sqrt{ \beta P_1P_2}}{1+(1-\rho^2) a^2P_2} \right) \\
R_1 \leq \frac{1}{2}\log \left(1+\frac{b^2P_1+\rho^2 P_2+2\rho b\sqrt{\beta P_1P_2}}{1+(1-\rho^2) P_2} \right) \\
R_2 \leq \frac{1}{2}\log \left(1+(1-\rho^2) P_2 \right) \\
0 \leq R_{2e} \leq R_2 \\
R_{2e}\leq\frac{1}{2}\log \left(1+(1-\rho^2) P_2 \right) \\
\hspace{1cm}-\frac{1}{2}\log \left(1+(1-\rho^2) a^2P_2 \right) \\
\end{array} \right\}
\end{split}
\end{equation}

%\begin{equation}\label{eq:gausscapaequi2}
%\begin{split}
%& \eufC = \bigcup_{0 \leq \beta \leq 1, 0 \leq \gamma \leq 1} \\
%& \left\{
%\begin{array}{l}
%(R_1,R_2,R_{2e}): \\
%R_1 \leq \frac{1}{2}\log \left(1+\frac{P_1+\bargamma a^2 P_2+2a\sqrt{\beta\bargamma P_1P_2}}{1+\gamma a^2P_2} \right) \\
%R_1 \leq \frac{1}{2}\log \left(1+\frac{b^2P_1+\bargamma P_2+2b\sqrt{\beta\bargamma P_1P_2}}{1+\gamma P_2} \right) \\
%R_2 \leq \frac{1}{2}\log \left(1+(1-\beta\bargamma) P_2 \right) \\
%R_1+R_2 \leq \frac{1}{2}\log \left(1+b^2P_1+P_2+2b \sqrt{\beta \bargamma P_1P_2} \right) \\
%R_1+R_2 \leq \frac{1}{2}\log \left(1+\frac{P_1+\bargamma a^2P_2+2a
%\sqrt{\beta \bargamma P_1P_2}}{1+\gamma a^2P_2} \right) \\
%\hspace{1.8cm}+\frac{1}{2}\log \left(1+\gamma P_2 \right) \\
%0 \leq R_{2e} \leq R_2 \\
%R_{2e}\leq\frac{1}{2}\log \left(1+\gamma P_2 \right)-\frac{1}{2}\log \left(1+\gamma a^2P_2 \right) \\
%\end{array} \right\}
%\end{split}
%\end{equation}
\end{theorem}
\qq

We note that the equivocation rate in this case can be positive.
This is because when $|a| < 1$, receiver 1's output $Y$ is
degraded with regard to receiver 2's output $Z$ if $X_1$ is given.
Hence receiver 2 may be able to receive some information that
receiver 1 cannot obtain. Note also, in
(\ref{eq:gausscapaequi2_deg}), if $a > 0,b > 0$, then $\beta=1$.

\begin{proof}(outline)
The achievability follows from \eqref{eq:capaequi degraded2} given
in Corollary \ref{cor:capaequi degraded2} by setting $V=X_2$ and
computing the mutual information terms with $(U,X_1,X_2)$ having
the following joint distribution:
\begin{equation}
X_1\sim {\cal N}(0,P_1),\; U=\rho\sqrt{\frac{\beta P_2}{P_1}}+U',
\;\;\text{and} \;\; X_2=U+X_2'
\end{equation}
where $ U'\sim{\cal N}(0,\bar{\beta}\rho^2P_2)$ and $X_2'\sim{\cal
N}(0,(1-\rho^2)P_2)$, and $X_1,U',X_2'$ are independent.

The converse follows by applying the bounds in the converse proof
for Corollary \ref{cor:capaequi degraded2} (see \cite{Liang07}) to
the Gaussian case and applying the entropy power inequality. The
proof also applies the degradedness property in this case, i.e.,
$Y$ is degraded with regard to $Z$ if $X_1$ is given. The details
of the proof are omitted in this paper and can be found in
\cite{Liang07}.
\end{proof}

\begin{figure}
\begin{center}
\psfrag{R1}[][][0.7]{$R_1$} \psfrag{R2}[][][0.7]{$R_2, R_{2,e}$}
\psfrag{a=0.5,R_2}[][][0.7]{$R_2,\;a=0.5$}
\psfrag{a=0.5,R_2e}[][][0.7]{$R_{2,e},a=0.5$}
\psfrag{a=8,R_2}[][][0.7]{$R_2,a=0.8$}
\psfrag{a=0.8,R_2e}[][][0.7]{$R_{2,e},a=0.8$}
\psfrag{a=1,R_2}[][][0.7]{$R_2,a=1$}
\psfrag{a=1,R_2e}[][][0.7]{$R_{2,e},a=1$}
\includegraphics[width=\linewidth]{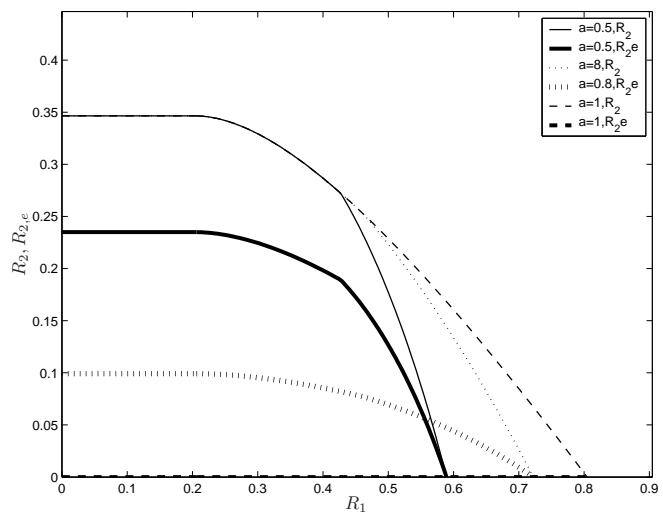}
\end{center}
\caption{The capacity region and the secrecy-capacity region of the
Gaussian cognitive interference channel for $P_1=P_2=1$, and
$b=1$.}\label{fig:low interference Gaussian IC}
\end{figure}

In Figure \ref{fig:low interference Gaussian IC} the capacity
region and the secrecy capacity region of the Gaussian cognitive
interference channel are shown for $P_1=P_2=1$, $b=1$ and
$a=0.5,0.8$ and $1$. It can be seen that as $a$ increases,
receiver 1 decodes more information about $W_2$ via rate
splitting. While this helps receiver 1 improve $R_1$ by
interference cancellation, it causes the equivocation rate
$R_{2e}$ to decrease due to more leakage of $W_2$ to receiver 1.
When $a=1$, receiver 1 decodes everything that receiver 2 decodes,
and hence $R_{2e}=0$, which is consistent with Theorem
\ref{th:gausscapaequi1}.

\section{Implication to Cognitive Interference Channels}\label{sec:implication}

Sections \ref{sec:mainresults} and \ref{sec:gauss} study the
cognitive interference channel with confidential messages. If we
do not consider the secrecy constraint, i.e., message $W_2$ need
not be confidential from receiver 1, the capacity-equivocation
region given in Theorem \ref{th:capaequi} reduces to the capacity
region of the corresponding cognitive interference channel without
secrecy constraints.
\begin{theorem}\label{th:capa}
The capacity region of the cognitive interference channel is given
by
%\begin{equation}\label{eq:capa}
%\begin{split}
%& C = \bigcup_{\displaystyle p(u,x_1,x_2)
%p(y,z|x_1,x_2)} \\
%& \left\{
%\begin{array}{l}
%(R_1,R_2): \\
%R_1 \ge 0, R_2 \ge 0 \\
%R_1 \leq \{I(U,X_1;Y),I(U,X_1;Z)\} \\
%R_2 \leq I(X_2;Z|X_1) \\
%R_1+R_2 \leq \min\{ I(U,X_1;Y), I(U,X_1;Z) \} \\
%\hspace{1.7cm}+I(X_2;Z|U,X_1) \\
%\end{array} \right\}
%\end{split}
%\end{equation}
%which is equivalent to the following region
\begin{equation}\label{eq:capa1}
\begin{split}
& C = \bigcup_{\displaystyle p(u,x_1,x_2)
p(y,z|x_1,x_2)} \\
& \left\{
\begin{array}{l}
(R_1,R_2): \\
R_1 \ge 0, R_2 \ge 0 \\
R_1 \leq I(U,X_1;Y) \\
R_2 \leq I(X_2;Z|X_1) \\
R_1+R_2 \leq \min\{ I(U,X_1;Y), I(U,X_1;Z) \} \\
\hspace{1.7cm}+I(X_2;Z|U,X_1) \\
\end{array} \right\}.
\end{split}
\end{equation}
\end{theorem}
\qq
\begin{proof}(outline)
From Corollary \ref{cor:capaequi1}, we deduce that the capacity region of
 the cognitive interference channel is given by (\ref{eq:capa1}) with an additional bound
$R_1\leq I(U,X_1;Z)$. This is done by setting $R_{2e}=0$ and because
the remaining bounds do not decrease if one sets $V=X_2$ due to the
Markov chain relationship $V \leftrightarrow (X_1,X_2)
\leftrightarrow (Y,Z)$. We further show that the bound $R_1\leq
I(U,X_1;Z)$ is, in fact, redundant.
\end{proof}

We note that the region \eqref{eq:capa1} was given as an
achievable rate region (inner bound on the capacity region) in
\cite[Theorem~5]{Jiang06}. The converse proof \cite{Liang07} that
we have given to show the more general result of Theorem
\ref{th:capaequi} provides a converse to establish that the region
\eqref{eq:capa1} is, in fact, the capacity region.

%We note that (\ref{eq:capa1}) is equivalent to the same region
%with $R_1$ satisfying an additional inequality $R_1 \leq
%I(U,X_1;Z)$.
We note that another achievable region for the cognitive
interference channel (without secrecy constraints) was reported in
\cite{Zhong07}, which is included within the larger achievable
region in \cite[Theorem~5]{Jiang06}.

The cognitive interference channel includes a few classical channels
as special cases.
\begin{remark}
The cognitive interference channel reduces to the broadcast channel
with degraded message sets studied in \cite{Korner772} if we set
$X_1=\phi$. Under this condition, it is easy to see that the
capacity region of cognitive interference channel given in Theorem
\ref{th:capa} reduces to the capacity region of the broadcast
channel with degraded message sets given in \cite[p.~360,
Theorem~4.1]{Csiszar78} that was shown to be equivalent to the
capacity region given in \cite{Korner772}.
\end{remark}

\begin{remark}
The capacity region of the cognitive interference channel reduces to
the capacity region of the multiple access channel with degraded
message sets given in \cite{Prelov84} (see also \cite{Slepian73}) if
we set $Y=Z$. In this case, the region given in \eqref{eq:capa1}
becomes
\begin{equation}
C = \bigcup_{\begin{array}{l} p(u,x_1,x_2)\\
p(z|x_1,x_2)\end{array}} \left\{
\begin{array}{l}
(R_1,R_2): \\
R_1 \ge 0, R_2 \ge 0 \\
R_1 \leq I(U,X_1;Z) \\
R_2 \leq I(X_2;Z|X_1) \\
R_1+R_2 \leq I(X_1,X_2;Z) \\
\end{array} \right\}
\end{equation}
It is easy to see that the preceding region is maximized by
setting $U=X_2$, and hence the first bound is not necessary. The
resulting region is the capacity region of the multiple access
channel with degraded message sets given in \cite{Prelov84} (see
also \cite{Slepian73}).
\end{remark}

\begin{corollary}
By setting $R_{2e}=0$, Theorems \ref{th:gausscapaequi1} and
\ref{th:gausscapaequi2} respectively reduce to the capacity
regions in the cases $|a|\geq 1$ and $|a|< 1$ for the Gaussian
cognitive interference channel without secrecy constraints, where
the cognitive receiver is required to reliably decode both
messages.
\end{corollary}

%%%%%%%%%%%%%%%%%%%%%%%%%%%%%%%%%%%%%%%%%%%%%%%%%%%%%%%%%%%%%%%%%%%%%%%%%%%%%%%%
\section{Conclusions}
In this paper we have presented a single-letter characterization
for the capacity-equivocation region of the cognitive interference
channel with confidential messages. The capacity-achieving random
scheme is based on superposition coding, rate-splitting and
stochastic encoding. We have further specialized the expression
for the capacity-equivocation region to several cases: (a)
\underline{perfect secrecy}, that is, the secrecy-capacity region;
(b) \underline{no secrecy constraints}, i.e., a new capacity
theorem for the cognitive interference channel; (c)
\underline{degraded channel} in which given the first channel
input, the observation available to the receiver that decodes both
messages is a degraded version of the observation available to the
eavesdropping receiver; and (d) \underline{degraded channel} in
which given the first channel input, the observation available to
the eavesdropping receiver is a degraded version of the
observation available to the other receiver. We have further
applied the results to the Gaussian cognitive interference channel
with confidential messages, which falls under cases (c) or (d),
and have explicitly characterized the capacity-equivocation region
in closed form expressions.

\end{document}